\documentclass{article}
\usepackage{graphicx}
\usepackage{cite}

\begin{document}

\title{Coherent States in Gravitational Quantum Mechanics}
\author{Pouria Pedram\thanks{p.pedram@srbiau.ac.ir}\\
{\small Department of Physics, Science and Research Branch, Islamic
Azad University, Tehran, Iran}}

\date{\today}
\maketitle \baselineskip 24pt

\begin{abstract}
We present the coherent states of the harmonic oscillator in the
framework of the generalized (gravitational) uncertainty principle
(GUP). This form of GUP is consistent with various theories of
quantum gravity such as string theory, loop quantum gravity, and
black-hole physics and implies a minimal measurable length. Using a
recently proposed formally self-adjoint representation, we find the
GUP-corrected Hamiltonian as a generator of the generalized
Heisenberg algebra. Then following Klauder's approach, we construct
exact coherent states and obtain the corresponding normalization
coefficients, weight functions, and probability distributions. We
find the entropy of the system and show that it decreases in the
presence of the minimal length. These results could shed light on
possible detectable Planck-scale effects within recent experimental
tests.
\end{abstract}


\textit{Pacs}: {04.60.Bc}

\section{Introduction}
The canonical quantization and the path integral quantization of
gravity are two well-known but old proposals which attempted to
quantize gravity and to unify the general relativity with the laws
of quantum mechanics. However, from field theoretical approach, the
theory of relativity is not renormalizable and results in the
ultraviolet divergences. Indeed, beyond the Planck energy scale, the
effects of gravity are so important which could lead to discreteness
of the very spacetime. This is due to the fact that when we probe
small distances with high energies, the spacetime structure will be
significantly disturbed by the gravitational effects. However, we
can solve the normalizability problem of gravity by introducing a
minimal measurable length as an effective cutoff in the ultraviolet
domain.

Various candidates of quantum gravity such as string theory, loop
quantum gravity and quantum geometry all agree on the existence of a
minimum observable length. In the language of the string theory, a
string cannot probe distances smaller than its length. Moreover,
some Gedanken experiments in black-hole physics and noncommutative
geometry imply a minimal length of the order of the Planck length
$\ell_{Pl}=\sqrt{\frac{G\hbar}{c^3}}\approx 10^{-35}m$ where $G$ is
Newton's gravitational constant. For instance, since in string
theory the mass of a string is proportional to its length, they
expand in size when probed at sufficiently high energies. This
suggests an additional momentum dependent uncertainty in the
position of a string in the form of \cite{1,12} ($\hbar=1$)
\begin{eqnarray}
\Delta X \geq \frac{1}{2\Delta P } +k\ell_{Pl}\Delta P,
\end{eqnarray}
where $k$ is a dimensionless constant. In order to incorporate the
idea of the minimal length into quantum mechanics, we need to change
the Heisenberg uncertainty principle to the so-called Generalized
Uncertainty Principle (GUP). The introduction of this idea has
attracted much attention in recent years and many papers have
appeared in the literature to address the effects of GUP on various
quantum mechanical systems
\cite{felder,13,14,15,16,r2,17,18,19,101,1011,102,103,104,105,106,7,61,Nouicer,pedram,pedram2,133,pedPhysA,pedramPLB2}.

In this paper, we are interested to find exact coherent states of
the harmonic oscillator in the framework of the generalized
commutation relation in the form $[X,P]=i\hbar(1+\beta P^{2})$ where
$\beta$ is the GUP parameter. Note that the problem of the
GUP-corrected harmonic oscillator is exactly solvable in the
momentum space and its exact energy eigenvalues and the
eigenfunctions are obtained in Refs.~\cite{Kempf,pedramPRD}.
Moreover, the perturbative construction of the corresponding
coherent states is discussed in Refs.~\cite{Nozari,Ghosh} to
first-order of the GUP parameter. Here, following Klauder's approach
and using the formally self-adjoint representation of the deformed
commutation relation, we take the Hamiltonian as a generator of the
generalized Heisenberg algebra and find the exact form of the
coherent states, weight functions, normalization coefficients, and
probability distributions. We show that the entropy of the system
reduces in the GUP scenario as a consequence of the minimal
observable length and we explain the physical reasoning behind this
phenomenon. The connection between our results and the recent
progresses in probing Planck-scale physics with quantum optics is
discussed finally.

\section{The Generalized Uncertainty Principle}
The Heisenberg uncertainty relation asserts that we can measure the
position and momentum of a particle separately with arbitrary
precision. So if there is a absolute minimal value for the results
of the measurements, the Heisenberg uncertainty relation should be
modified. Here we consider a generalized uncertainty principle which
implies a minimum observable length
\begin{eqnarray}\label{gup}
\Delta X \Delta P \geq \frac{\hbar}{2} \left( 1 +\beta \left[(\Delta
P)^2+\langle P\rangle^2 \right]\right),
\end{eqnarray}
where $\beta$ is the GUP parameter. We also have
$\beta=\beta_0/(M_{Pl} c)^2$ where $M_{Pl}$ is the Planck mass and
$\beta_0$ is of the order of unity. GUPs of this type also arise
from polymer quantization \cite{p1,p2,p3}. Note that the deviation
from the Heisenberg picture takes place in the high energy domain
where the effects of gravity would be significant. Thus, for the
energies much lower than the Planck energy $M_{Pl} c^2\sim 10^{19}$
GeV, we recover the well-known Heisenberg uncertainty relation. It
is easy to check that the above inequality relation (\ref{gup})
implies the existence of an absolute minimum observable length given
by $(\Delta X)_{min}=\hbar\sqrt{\beta}$. In the context of string
theory, we can interpret this length as the length of the string and
conclude that the string's length is proportional to the square root
of the GUP parameter. In one spatial dimension, the above
uncertainty relation can be obtained from the following deformed
commutation relation
\begin{eqnarray}\label{gupc}
[X,P]=i\hbar(1+\beta P^2).
\end{eqnarray}
As it is recently suggested in
Ref.~\cite{pedramPRD,pedramPLB}, we can write $X$ and $P$ in terms
of ordinary position and momentum operator as
\begin{eqnarray}\label{k1}
X &=& x,\\ P &=&
\frac{\tan\left(\sqrt{\beta}p\right)}{\sqrt{\beta}},\label{k2}
\end{eqnarray}
where $[x,p]=i\hbar$ and $X$ and $P$ are symmetric operators on the
dense domain $S_{\infty}$ with respect to the following scalar
product:
\begin{eqnarray}\label{scalar}
\langle\psi|\phi\rangle=\int_{-\frac{\pi}{2\sqrt{\beta}}}^{+\frac{\pi}{2\sqrt{\beta}}}\mathrm{d}p\,\psi^{*}(p)\phi(p).
\end{eqnarray}
With this definition, the commutation relation (\ref{gupc}) is
exactly satisfied. In this representation, the completeness relation
and scalar product can be written as
\begin{eqnarray}\label{comp1}
\langle p'|p\rangle= \delta(p-p'),\\
\int_{-\frac{\pi}{2\sqrt{\beta}}}^{+\frac{\pi}{2\sqrt{\beta}}}\mathrm{d}
p\, |p\rangle\langle p|=1.\label{comp2}
\end{eqnarray}
Also the eigenfunctions of the position operator in momentum space
is given by the solutions of the eigenvalue equation
\begin{eqnarray}
X\,u_x(p)=x\,u_x(p),
\end{eqnarray}
where $u_x(p)=\langle p|x\rangle$. The normalized solution is
\begin{eqnarray}
u_x(p)=\sqrt{\frac{\sqrt{\beta}}{\pi}}\exp\left({-i\frac{ p}{\hbar}}
x\right),
\end{eqnarray}
which can be used to check the scalar product relation
(\ref{comp1}). Now using (\ref{comp2}) we find the wave function in
coordinate space as
\begin{eqnarray}
\psi(x)=\sqrt{\frac{\sqrt{\beta}}{\pi}}
\int_{-\frac{\pi}{2\sqrt{\beta}}}^{+\frac{\pi}{2\sqrt{\beta}}}
e^{\frac{i px}{\hbar}}\phi(p)\mathrm{d} p.
\end{eqnarray}

Note that this representation is equivalent with the seminal
proposal by Kempf, Mangano and Mann (KMM) \cite{Kempf}
\begin{eqnarray}\label{x0p01}
X &=& (1+\beta p^2)x,\\
P &=& p,\label{x0p02}
\end{eqnarray}
through the following canonical transformation:
\begin{eqnarray}
X&\longrightarrow&\left[1+\arctan^2\left(\sqrt{\beta}P\right)\right]X,\\
P&\longrightarrow&\arctan\left(\sqrt{\beta}P\right)/\sqrt{\beta},
\end{eqnarray}
which transforms Eqs.~(\ref{x0p01}) and (\ref{x0p02}) to
Eqs.~(\ref{k1}) and (\ref{k2}) subjected to Eq.~(\ref{gupc}). This
representation (\ref{k1}-\ref{k2}) preserves the ordinary nature of
the position operator and only affects the kinetic part of the
Hamiltonian. Similar to KMM representation the momentum operator $P$
is self-adjoint, while the position operator $X$ is merely
symmetric. This is due to the fact that the domain of $X^\dagger$ is
much larger than the domain of $X$. However, this representation is
\emph{formally} self-adjoint, i.e., $A=A^\dagger$ for $A\in\{X,P\}$
(see \cite{pedramPRD} for details). Now $P$ and $p$ can be
interpreted as follows: $p$ is the momentum operator at low energies
($p=-i\hbar
\partial/\partial{x}$) and $P$ is the momentum operator at
high energies. Obviously, this procedure affects all Hamiltonians in
the quantum mechanics, namely
\begin{eqnarray}
H=\frac{P^2}{2m} + V(X),
\end{eqnarray}
where using Eqs.~(\ref{k1}) and (\ref{k2}) can be rewritten as
\begin{eqnarray}
H= \frac{\tan^2\left(\sqrt{\beta}p\right)}{2\beta m}+ V(x).
\end{eqnarray}
Since this Hamiltonian is formally self-adjoint ($H^\dagger=H$), we
can use the general scheme of the coherent states to find the
desirable solutions, whereas this property is absent in other
representations \cite{Kempf,r1}. In the quantum domain, this
Hamiltonian results in the following generalized Schr\"odinger
equation in coordinate space
\begin{eqnarray}
-\frac{\hbar^2}{2m}\frac{\partial^2\psi(x)}{\partial x^2}+
\sum_{n=3}^\infty  \frac{2^{2n} (2^{2n}-1)(2n-1)
B_{2n}\hbar^{2(n-1)}\beta^{n-2}}{2m(2n)!}
\frac{\partial^{2(n-1)}\psi(x)}{\partial
x^{2(n-1)}}+V(x)\psi(x)=E\,\psi(x),\hspace{1cm}
\end{eqnarray}
where $B_n$ is the $n$th Bernoulli number.

\section{GUP and the Harmonic oscillator}
For the harmonic oscillator, because of the quadratic form of the
potential $V(x)=(1/2)\,m\omega^2x^2$, we obtain a second-order
differential equation in the momentum space, namely \cite{pedramPRD}
\begin{eqnarray}\label{HamilSHO}
-\frac{\partial^2\phi(p)}{\partial
p^2}+\frac{\tan^2\left(\sqrt{\gamma}p\right)}{\gamma}\phi(p)=\epsilon\,\phi(p),
\end{eqnarray}
where $p\rightarrow\sqrt{m\hbar\omega}\,p$,
$\gamma=m\hbar\omega\beta$, and
$\epsilon=\displaystyle\frac{2E}{\hbar\omega}$. In terms  of the new
variable $y=\sqrt{\gamma}p$, it reads
\begin{eqnarray}
\left[-\frac{\partial^2}{\partial
y^2}+\nu(\nu-1)\tan^2(y)-\bar\epsilon(\nu)\right]\phi(y;\nu)=0,
\end{eqnarray}
where by definition
\begin{eqnarray}
\nu=\frac{1}{2}\left(1+\sqrt{1+\frac{4}{\gamma^2}}\right),\hspace{2cm}\bar\epsilon=\frac{\epsilon}{\gamma},
\end{eqnarray}
and the boundary condition is
\begin{eqnarray}
\phi(y;\nu)\bigg|_{y=\pm\pi/2}=0.
\end{eqnarray}
The above differential equation is exactly solvable and the
eigenfunctions can be obtained in terms of the Gauss hypergeometric
functions \cite{pedramPRD}
\begin{eqnarray}\label{solSHO1}
\phi_{2k}(p\,;\gamma)&=&{\cal
A}_k(\nu)\left[\cos(\sqrt{\gamma}p)\right]^{\left(1+\sqrt{1+\frac{4}{\gamma^2}}\right)/2}\nonumber\\
&&\times\,{}_2F_1\left(-k,\nu+k; \nu+ \frac{1}{2};
\cos^2(\sqrt{\gamma}p)\right),\hspace{.5cm}k=0,1,2,\ldots,
\end{eqnarray}
for even states and
\begin{eqnarray}\label{solSHO2}
\phi_{2k+1}(p\,;\gamma)&=&{\cal B}_k(\nu)\sin(\sqrt{\gamma}p)
\left[\cos(\sqrt{\gamma}p)\right]^{\left(1+\sqrt{1+\frac{4}{\gamma^2}}\right)/2}\nonumber
\\ &&\times\,{}_2F_1\left(-k,\nu+k+1; \nu+ \frac{1}{2};
\cos^2(\sqrt{\gamma}p)\right),\hspace{.5cm}k=0,1,2,\ldots,\hspace{1cm}
\end{eqnarray}
for odd states. Moreover, the exact GUP-corrected energy spectrum is
given by
\begin{eqnarray}\label{E-exact}
\epsilon_n=\left(2n+1\right)\left(\sqrt{1+\frac{\gamma^2}{4}}+\frac{\gamma}{2}\right)+\gamma
n^2,\hspace{1cm}n=0,1,2,\ldots\,.
\end{eqnarray}
The generalization of this problem to arbitrary dimension is also
discussed in Ref.~\cite{r1}.

\section{The Generalized Coherent States}
Coherent states were originally introduced by Schr\"odinger in 1926
\cite{Sch} have applications in many areas of physics
\cite{Tref1,Tref2,Tref3,Tref4,Tref5,Tref6}. To construct coherent
states, we follow Klauder's approach \cite{Klauder1,Klauder2} and
use the version of the generalized Heisenberg algebra (GHA)
\cite{f1,f2,f3,11,ref} given in Refs.~\cite{11,ref}. This version of
GHA consists of the generators $J_0$, $A$, and $A^\dagger$ that
satisfy \cite{11}
\begin{eqnarray}
J_0A^\dagger &=& A^\dagger f(J_0),\label{g1}\\
AJ_0 &=& f(J_0)A,\label{g2}\\
\left[A^{\dagger},A\right] &=& J_0-f(J_0).\label{g3}
\end{eqnarray}
Here $A=(A^\dagger)^\dagger$ and $J_0=J_0^\dagger$ is the
Hamiltonian of the system. Moreover, $f(J_0)$ is an analytic
function of $J_0$ and is called the characteristic function of the
algebra. These generators also satisfy
\begin{eqnarray}
J_0|m\rangle &=& \alpha_m|m\rangle,\\
A^\dagger|m\rangle &=& N_m|m+1\rangle,\\
A|m\rangle &=& N_{m-1}|m-1\rangle,
\end{eqnarray}
where $N_m^2=\alpha_{m+1}-\alpha_0$. For instance, for the linear
function $f(x)=x+1$, we obtain the harmonic oscillator algebra, and
for $f(x)=qx+1$, the algebra in Eqs.~(\ref{g1})-(\ref{g3}) becomes
the deformed Heisenberg algebra \cite{11}. It is shown that the
quantum systems having the energy spectrum
\begin{eqnarray}
\epsilon_{n+1}=f(\epsilon_{n}),
\end{eqnarray}
where $\epsilon_{n}$ and $\epsilon_{n+1}$ are successive energy
levels and $f(x)$ is a distinct function for each physical system,
are described by these generalized Heisenberg algebras. Also, This
function is exactly the same function that appears in the
construction of the algebra, i.e., the characteristic function of
the algebra. In the algebraic langauge, $J_0$ is the Hamiltonian,
$A$ is the annihilation operator, and $A^\dagger$ is the creation
operator. These operators are related to the Casimir operator of the
GHA through the relation
\begin{eqnarray}
C=A^\dagger A-J_0=A A^\dagger-f(J_0).
\end{eqnarray}
Now the coherent states are given by \cite{ref}
\begin{eqnarray}\label{CS}
|z\rangle=N(z)\sum_{n=0}^{\infty}\frac{z^n}{N_{n-1}!}|n\rangle,
\end{eqnarray}
where $A|z\rangle=z|z\rangle$, $N(z)$ is the normalization
coefficient, by definition $N_{n}!\equiv N_0N_1\cdots N_n$, and by
consistency $N_{-1}!\equiv1$. Note that Klauder's coherent states
should satisfy the following minimal set of conditions \cite{ref}:
\begin{enumerate}
  \item normalizability condition
  \begin{eqnarray}
\langle z|z\rangle=1.
\end{eqnarray}
  \item continuity in the label
\begin{eqnarray}
|z-z'|\rightarrow0,\hspace{1cm}|| |z\rangle-|z'\rangle
||\rightarrow0.
\end{eqnarray}
  \item completeness relation
  \begin{eqnarray}
\int \mathrm{d}^2z\, w(z)\, |z\rangle\langle z|=1.
\end{eqnarray}
\end{enumerate}

\begin{figure}
\begin{center}
\includegraphics[width=8cm]{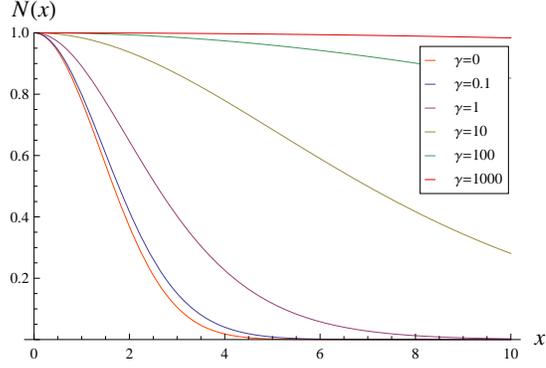}
\caption{\label{fig1}Normalization function for the GUP-corrected
harmonic oscillator and for $\gamma=\{0,0.1,1,10,100,1000\}$.}
\end{center}
\end{figure}
For the harmonic oscillator with the minimal length uncertainty,
using Eq.~(\ref{E-exact}) we obtain
\begin{eqnarray}
\epsilon_{n+1}&=&\epsilon_{n}+2\gamma
(n+1)+\sqrt{4+\gamma^2},\nonumber\\
&=&\epsilon_{n}+2\sqrt{\gamma\epsilon_{n}+1}+\gamma.
\end{eqnarray}
Thus, the characteristic function reads
\begin{eqnarray}
f(x)=x+2\sqrt{\gamma x+1}+\gamma.
\end{eqnarray}
Since the above algebraic formalism implies $\alpha_n=\epsilon_{n}$,
we find
\begin{eqnarray}
N^2_{n-1}&=&\alpha_n-\alpha_0=\gamma\left[
n^2+\left(\sqrt{1+\frac{4}{\gamma^2}}+1\right)n\right],\nonumber\\
&=&\gamma\left( n+2\nu \right)n,
\end{eqnarray}
which results in
\begin{eqnarray}\label{N!}
N_{n-1}!=\gamma^{n/2}\sqrt{\frac{n!(n+2\nu)!}{(2\nu)!}}.
\end{eqnarray}
So the coherent states given in Eq.~(\ref{CS}) can be written as
\begin{eqnarray}
|z\rangle=N(|z|)\sqrt{(2\nu)!}\sum_{n=0}^{\infty}\frac{\gamma^{-n/2}z^n}{\sqrt{n!}\sqrt{(n+2\nu)!}}|n\rangle.
\end{eqnarray}
The normalizability condition is given by
\begin{eqnarray}
1=N^2(|z|)(2\nu)!\sum_{n=0}^{\infty}\frac{\gamma^{-n}|z|^{2n}}{n!(n+2\nu)!}.
\end{eqnarray}
Since we have
\begin{eqnarray}
\sum_{n=0}^{\infty}\frac{\gamma^{-n}|z|^{2n}}{n!(n+2\nu)!}=\frac{I_{2\nu}\left(\frac{2|z|}{\sqrt{\gamma}}\right)}{\gamma^{-\nu}|z|^{2\nu}},
\end{eqnarray}
where $I_p(z)$ is the modified Bessel function of the first kind of
order $p$ and $0\leq |z|<\infty$, the normalizability condition
reads
\begin{eqnarray}
N^2(|z|)=\frac{\gamma^{-\nu}|z|^{2\nu}}{(2\nu)!\,I_{2\nu}\left(\frac{2|z|}{\sqrt{\gamma}}\right)}.
\end{eqnarray}
In Fig.~\ref{fig1} we have depicted $N(x)$ for various values of
$\gamma$. Note that for $\gamma\rightarrow0$ (harmonic oscillator
without GUP), $N(x)$ goes to $e^{-x^2/4}$ and for
$\gamma\rightarrow\infty$ ($\nu\rightarrow1$) (particle in a box),
it tends to
$\displaystyle\frac{x^2/\gamma}{2I_2(2x/\sqrt{\gamma})}$.

\begin{figure}
\begin{center}
\includegraphics[width=8cm]{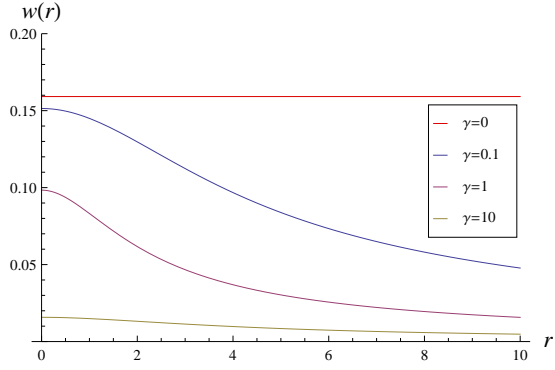}
\caption{\label{fig2}Weight function for the GUP-corrected harmonic
oscillator and for $\gamma=\{0,0.1,1,10\}$.}
\end{center}
\end{figure}
\begin{figure}
\begin{center}
\includegraphics[width=8cm]{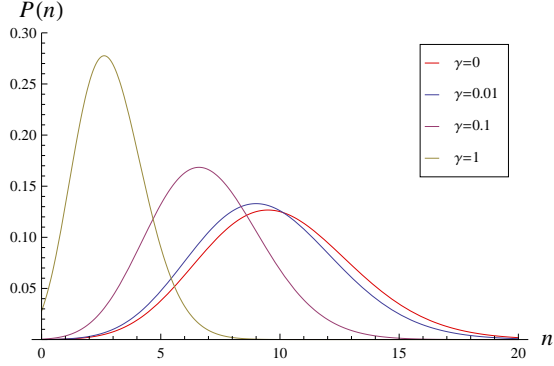}
\caption{\label{fig3}The probability distribution for the
GUP-corrected harmonic oscillator. We set $\lambda=20$
($\overline{n}=10$ for $\beta=0$) and $\gamma=\{0,0.01,0.1,1\}$.}
\end{center}
\end{figure}

To satisfy the completeness relation, we need to find the adequate
weight function $w(r)$, $z=re^{i\theta}$, implying the equality
\begin{eqnarray}
2\pi\sum_{n=0}^{\infty}|n\rangle\langle
n|\frac{(2\nu)!\,\gamma^{-n}}{n!(n+2\nu)!}\int_0^{\infty}\mathrm{d}r\,
N^2(r)w(r) r^{2n+1}=1.
\end{eqnarray}
If we take $x=r^2$, we obtain
\begin{eqnarray}
\pi(2\nu)!\sum_{n=0}^{\infty}|n\rangle\langle
n|\frac{2\gamma^{-n-\nu-1}}{n!(n+2\nu)!}\int_0^{\infty}\mathrm{d}x\,
\frac{\gamma^{\nu+1}N^2(\sqrt{ x})w(\sqrt{ x})}{2x^\nu}
x^{n+\nu}=1.\label{wi}
\end{eqnarray}
So by taking
\begin{eqnarray}
\frac{\pi(2\nu)!\,\gamma^{\nu+1}N^2(\sqrt{ x})w(\sqrt{
x})}{2x^\nu}=K_{2\nu}\left(2\sqrt{\frac{x}{\gamma}}\right),
\end{eqnarray}
where $K_p(x)$ is the modified Bessel function of the second kind of
order $p$ and using
\begin{eqnarray}
\int_0^{\infty}\mathrm{d}x\,K_{2\nu}\left(2\sqrt{\frac{x}{\gamma}}\right)
x^{n+\nu}=\frac{1}{2}\gamma^{n+\nu+1}n!(n+2\nu)!,
\end{eqnarray}
Eq.~(\ref{wi}) is satisfied which gives the following weight
function
\begin{eqnarray}
w(\sqrt{x})=\frac{2x^\nu
K_{2\nu}(2\sqrt{x})}{\pi(2\nu)!\,\gamma^{\nu+1}N^2(\sqrt{\gamma
x})},
\end{eqnarray}
where can be finally expressed as
\begin{eqnarray}
w(r)=\frac{2}{\pi
\gamma}I_{2\nu}\left(\frac{2r}{\sqrt{\gamma}}\right)K_{2\nu}\left(\frac{2r}{\sqrt{\gamma}}\right).
\end{eqnarray}
In Fig.~\ref{fig2}, the behavior of the weight function is shown for
$\gamma=\{0,0.1,1,10\}$. As we have expected, for
$\gamma\rightarrow0$ (harmonic oscillator without GUP), $w(r)$ tends
to $\displaystyle\frac{1}{2\pi}$ and for $\gamma\rightarrow\infty$
($\nu\rightarrow1$) (particle in a box), it goes to
$\displaystyle\frac{2}{\pi
\gamma}I_{2}\left(\frac{2r}{\sqrt{\gamma}}\right)K_{2}\left(\frac{2r}{\sqrt{\gamma}}\right)$.
It is worth to mention that a potential application of our results
is in quantum optics. Indeed the coherent states as the states of
the light field can be used to approximately describe the output of
a single-frequency laser well above the laser threshold. In the
absence of GUP, the probability of detecting $n$ photons is given by
Poisson distribution, namely
\begin{eqnarray}\label{Poisson}
P(n;\lambda)=|\langle
n|z\rangle|^2=\frac{e^{-\lambda/2}}{n!}\left(\frac{\lambda}{2}\right)^n,
\end{eqnarray}
where $\lambda=|z|^2=\langle z|A^\dagger A|z\rangle$ and $\lambda/2$
is the average photon number $\overline{n}$ in a coherent state for
$\gamma=0$ ($\epsilon_n=2n+1$).\footnote{In terms of $\overline{n}$
we have $P(n)=\frac{e^{-\bar{n}}\,\bar{n}^n}{n!}$.} In the presence
of the minimal length, the probability distribution is
\begin{eqnarray}
P(n;\lambda,\gamma)&=&|\langle
n|z\rangle|^2=N^2(\sqrt{\lambda})\frac{\lambda^n}{(N_{n-1}!)^2},\label{P1} \\
&=&\frac{\gamma^{-n-\nu}}{I_{2\nu}\left(2\sqrt{\frac{\lambda}{\gamma}}\right)}\frac{\lambda^{n+\nu}}{n!(n+2\nu)!},\label{P2}
\end{eqnarray}
which encodes the effects of GUP on the harmonic oscillator
statistics and satisfies
\begin{eqnarray}
\sum_{n=0}^\infty P(n;\lambda,\gamma)=1.
\end{eqnarray}
Also we have
\begin{eqnarray}\label{P-sub}
\frac{\lambda}{\gamma}=\overline{n^2}+2\nu\overline{n},
\end{eqnarray}
which results in
\begin{eqnarray}\label{P-2}
P(n;\overline{n},\nu)=\frac{\left(\overline{n^2}+2\nu\overline{n}\right)^{n+\nu}}{I_{2\nu}\left(2\sqrt{\overline{n^2}+2\nu\overline{n}}\right)\,n!(n+2\nu)!}.
\end{eqnarray}
For $\gamma\rightarrow0$, we have
$N^2(\sqrt{\lambda})=e^{-\lambda/2}$ (see Fig.~\ref{fig1}) and using
\begin{eqnarray}
\lim_{\nu\rightarrow\infty}\frac{\nu^{-n}\,(n+2\nu)!}{(2\nu)!}=2^n,
\end{eqnarray}
Eq.~(\ref{N!}) reads
\begin{eqnarray}
N_{n-1}!\simeq\sqrt{2^n\,n!}\,\,.
\end{eqnarray}
So Eq.~(\ref{P1}) gives the Poisson distribution (\ref{Poisson}) at
this limit. For $\gamma\rightarrow\infty$ ($\nu\rightarrow1$),
Eq.~(\ref{P2}) is expressed as
\begin{eqnarray}
P(n;\lambda,\gamma\rightarrow\infty)=\frac{\gamma^{-n-1}}{I_{2}\left(2\sqrt{\frac{\lambda}{\gamma}}\right)}\frac{\lambda^{n+1}}{n!(n+2)!},
\end{eqnarray}
where
$\displaystyle\frac{\lambda}{\gamma}\simeq\overline{n^2}+2\overline{n}$.
Fig.~\ref{fig3} shows the schematic behavior of the probability
distribution for the GUP-corrected harmonic oscillator and for
various values of the GUP parameter. As the figure shows, $n_{max}$
which gives the maximum probability, decreases as $\gamma$ increases
and is given by the root of the following equation:
\begin{eqnarray}
H_{n_{max}}+H_{n_{max}}^{(2\nu)}+\ln\frac{\gamma}{\lambda}-2\xi=0,
\end{eqnarray}
where $H_p$ is $p$th harmonic number, $H_p^{(r)}$  is $p$th harmonic
number of order $r$, and $\xi$ is Euler's constant. Note that
Eqs.~(\ref{P2}) and (\ref{P-2}) define the probability of detecting
$n$ photons in a laser beam subject to (\ref{P-sub}) which is a
result of the deformed generalized uncertainty relation (expected to
be significant at very high energies where gravity is considerable).
Therefore, for a fixed $\lambda$, the averaged number of photons
decreases as the GUP parameter increases.

\begin{figure}
\begin{center}
\includegraphics[width=8cm]{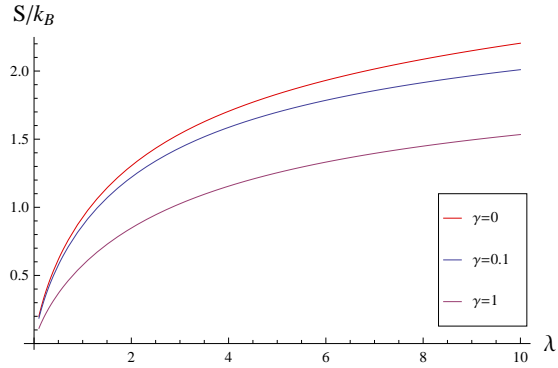}
\caption{\label{fig4}The entropy of the GUP-corrected harmonic
oscillator for $\gamma=\{0,0.1,1\}$.}
\end{center}
\end{figure}

Now we can define the entropy of this system as the logarithmic
measure of the density of states:
\begin{eqnarray}
S(\lambda;\gamma)=-k_B\sum_{n=0}^\infty P(n;\lambda,\gamma)\,\ln
P(n;\lambda,\gamma),
\end{eqnarray}
where $k_B$ is the Boltzmann constant. In Fig.~\ref{fig4}, we have
depicted the entropy of the GUP-corrected harmonic oscillator for
$\gamma=\{0,0.1,1\}$. As the figure shows, for fixed $\lambda$, the
entropy decreases as $\gamma$ increases and tends to the Poisson
entropy for $\gamma\rightarrow0$
\begin{eqnarray}
S(\lambda;0)=k_B\left[\frac{\lambda}{2}\left(1-\ln
\frac{\lambda}{2}\right)+e^{-\lambda/2}\sum_{n=0}^{\infty}\frac{\left(\frac{\lambda}{2}\right)^n
\ln n!}{n!}\right].
\end{eqnarray}
The reason behind this behavior can be understood from the GUP
commutation relation (\ref{gupc}). Indeed, we can consider the right
hand side of this equation effectively as the GUP-corrected Planck's
constant which is always greater than $\hbar$. Therefore, in the
language of the statistical mechanics, the size of the unit cell in
the phase space increases and consequently the number of the
accessible states and the entropy of the system decrease with
respect to the absence of GUP.

As we state before, a potential application of our calculations is
in quantum optics. But the question is: could the relation between
quantum optics and Plank-scale uncertainty relations have some
detectable effects? To answer this question we should mention that
in recent years various approaches are developed to test the effects
of quantum gravity and to explore possible quantum gravitational
phenomena. These attempts range from astronomical observations
\cite{laser3,laser3-2} to table-top experiments \cite{laser4}.
Amelino-Camelia and Lammerzahl proposed some laser interferometric
setups to explain puzzling observations of ultrahigh energy cosmic
rays in the context of quantum gravity modified laws of particle
propagation \cite{laser1}. The implications of high intensity Laser
projects for quantum gravity phenomenology are also discussed by
Magueijo based on deformed special relativity \cite{laser2}.

Recently, Pikovski \emph{et al.} have introduced a scheme to
experimentally test the existence of a minimal length scale as a
modification of the Heisenberg uncertainty relation for various GUP
scenarios in the context of quantum optics \cite{laser4}. They
utilized quantum optical control to probe possible deviations from
the quantum commutation relation at the Planck scale and showed that
their scheme is within reach of current technology. In fact, the
idea is direct measurement of the canonical commutator of a massive
object using a quantum optical ancillary system that produces
nonlinear enhancement without need for Planck-scale accuracy of
position measurements. So the possible Planck-scale commutator
deformations can be observed with very high accuracy by optical
interferometric techniques that are within experimental reach. These
experimental progresses support the implication of our calculations
which could shed light on possible detectable Planck-scale effects
with quantum optics.

\section{Conclusions}
In this paper, we have studied the construction of the coherent
states of the harmonic oscillator in the context of the generalized
uncertainty principle which implies a minimal length uncertainty
proportional to the Planck length. Following Klauder's approach, we
constructed the generalized Heisenberg algebra where the Hamiltonian
was its formally self-adjoint generator. Then, after finding the
characteristic function of the algebra, we obtained the exact
expression for the coherent states, weight functions, normalization
coefficients, and probability distributions and studied their
behavior in terms of the GUP parameter. We showed that because of
the gravitational uncertainty relation the ordinary quantum
description of laser light should be modified and the Poisson
probability distribution will not be exactly preserved. Also, the
entropy of the system decreased in the presence of the minimal
length uncertainty due to the increase of the size of the unit cell
in the phase space. Finally, we indicated that recent progresses to
experimentally test the existence of a minimal length scale could
reveal some evidence of possible quantum gravitational effects.

\section*{Acknowledgments}
I am very grateful to Kourosh Nozari for fruitful discussions and
suggestions and for a critical reading of the manuscript.

\end{document}